\documentclass[12pt]{article}
\usepackage{amsfonts}
\begin{document}
\def\a{\alpha}\def\b{\beta}\def\g{\gamma}\def\d{\delta}\def\e{\epsilon }
\def\k{\kappa}\def\l{\lambda}\def\L{\Lambda}\def\s{\sigma}\def\S{\Sigma}
\def\Th{\Theta}\def\th{\theta}\def\om{\omega}\def\Om{\Omega}\def\G{\Gamma}
\def\y{\vartheta}\def\m{\mu}\def\n{\nu}
\def\ws{worldsheet}
\def\susy{supersymmetry}
\def\ts{target superspace}
\def\ks{$\k$-symmetry}

\font\mybb=msbm10 at 12pt
\def\bb#1{\hbox{\mybb#1}}
\renewcommand{\theequation}{\arabic{section}.\arabic{equation}}

\title{\bf Extended supersymmetry in massless conformal higher spin theory}

\bigskip
\author{Igor A. Bandos
$^{\dagger\ddagger} $, Jos\'e A. de Azc\'arraga$^{\ast}$ and Carlos Meliveo$^{\dagger}$,
\\ $^{\dagger}$ \small\it Department of
Theoretical Physics, the University of the Basque Country UPV/EHU,  \\
\small\it P.O. Box 644, 48080 Bilbao, Spain
 \\  \small\it $^{\ddagger} $
IKERBASQUE, Basque Foundation for Science, 48011, Bilbao, Spain \\
 { $^{\ast}$ \small\it Departamento de F\'{\i}sica Te\'orica and IFIC (CSIC-UVEG),
 46100-Burjassot (Valencia), Spain}
}

\date{\small{June 26, 2011}}

 \maketitle

\begin{abstract}
{\small
We propose superfield equations in tensorial ${\cal N}$-extended superspaces
to describe the ${\cal N}=2,4,8$ supersymmetric generalizations of free conformal
higher spin theories. These can be obtained by quantizing a superparticle model
in ${\cal N}$-extended tensorial superspace.  The  ${\cal N}$-extended higher
spin supermultiplets just contain scalar and `spinor' fields in tensorial space
so that, in contrast with the standard (super)space approach, no nontrivial generalizations
of the Maxwell or Einstein equations to tensorial space appear when ${\cal N}>2$.
For ${\cal N}=4, 8$, the higher spin-tensorial components of the extended tensorial
superfields are expressed through additional scalar and spinor fields in tensorial
space which obey the same free higher spin equations, but that are axion-like in the
sense that they possess Peccei-Quinn-like symmetries. }

\end{abstract}


\maketitle

\def\theequation{\arabic{section}.\arabic{equation}}
\setcounter{equation}0
\section{Introduction}

It is known \cite{Frons:85,BLS99,V01s,V01c,Dima,O+Misha03,BPST04,BBdAST05}
that the $D=4$ free conformal higher spin equations can be formulated as
a field theory on a ten-dimensional tensorial space, $\Sigma^{(10|0)}$,
parametrized by 10 bosonic coordinates $(x^m,y^{mn})$,
\begin{equation}
\label{x} X^{\alpha\beta}=X^{\beta\alpha}= {1\over 4}x^m\gamma_m^{\alpha\beta} + {1\over 8}
y^{mn}\gamma_{mn}^{\alpha\beta}\;  \qquad \alpha , \beta =1,2,3,4\; , \quad m,n=0,1,2,3 \; .
\quad
\end{equation}
and in ${\cal N}=1$ tensorial superspace $\Sigma^{(10|4)}$ with coordinates
$(X^{\alpha\beta},\theta^\alpha)$.

The bosonic tensorial space (\ref{x}) was proposed as a natural
basis to build $D$=4 conformal higher spin theories in \cite{Frons:85}.
More general tensorial spaces of dimension ${n(n+1)\over 2}$ are
introduced by means of a symmetric $n\times n$ coordinates
matrix $X^{\alpha\beta}$ $(\alpha,\beta=1,\dots,n)$
which, for even $n=4,8,16$, determine tensorial enlargements of the standard
spacetimes of dimension $D$=4,6,10 ($D={n \over 2}+2)$. Adding
$n$ fermionic coordinates $\theta^\alpha$ one obtains the ${\cal N}=1$
tensorial superspaces $\Sigma^{({n(n+1)\over 2}|n)}$,
\begin{equation}
\label{x+th} \Sigma^{({n(n+1)\over 2}|n)}\; : \qquad  Z^{\cal M}:= (X^{\alpha\beta}\, , \,
\theta^\alpha) \; , \qquad \cases{\alpha,\beta =1,...,n , \cr
X^{\alpha\beta}=X^{\beta\alpha}\;  ;  } \qquad
\end{equation}
 rigid tensorial superspaces $\Sigma^{({n(n+1)\over 2}|n)}$
have a supergroup structure. Taking $n$ even is not a restriction if we
think of the underlying spinorial origin of the $\alpha$ indices;
furthermore this motivates also the restriction to $n=2^k=2,4,8,16,...$
assuming that $\theta^\alpha$ are spinors. Although the fermionic coordinates
in $\Sigma^{({n(n+1)\over 2}|n)}$ are usually assumed to be real, in the $n=4$, $D=4$
case it is convenient to consider $\theta^\alpha$ as a Majorana spinor
in the Weyl realization of the Dirac matrices so that
$\theta^\alpha=\left(\theta^A, \bar{\theta}_{\dot{A}}\right)$.

For $n=2$ the spin-tensorial coordinates $X^{\alpha\beta}$ are expressed in terms
of the 3-vector spacetime coordinates,
$X^{\alpha\beta}\propto x^a\tilde{\gamma}^{\alpha\beta}_a$ so that $\Sigma^{(3|2)}$
is just the standard $D=3$, ${\cal N}=1$ superspace. The case $n=32$ gives the tensorial
extension of $D$=11 superspace $\Sigma^{(528|32)}$, relevant in the context
of the BPS preon hypothesis \cite{BPS01,BdAIPV:03} and also in the analysis of
the hidden gauge structure of $D$=11 supergravity \cite{D'Au-Fre:82,BdAPV05}.
In this paper we shall restrict ourselves to the $n$=4,8,16 cases that are used
to describe massless conformal higher spin theories in $D=4,6,10$. Almost all our
equations (all but those in Sec.~4.2.1) will be valid for all these dimensions,
although we shall make special emphasis on the $n=4$ case corresponding to $D=4$.

The first mechanical model in ${\cal N}$=1, $D$=4 tensorial superspace
$\Sigma^{(10|4)}$ and in its higher dimensional generalizations
$\Sigma^{({n(n+1)\over 2}|n)}$ with $n>4$ was proposed in \cite{BL98},
where it was noticed that the ground state of such a superparticle
model describes BPS states preserving all but one supersymmetry.  The possible
`constituent' r\^ole of such states in string/M-theory was introduced and discussed in
general in \cite{BPS01}, where they were called `BPS preons' (see further \cite{BdAIPV:03}).
Thus, from this viewpoint, the superparticle in \cite{BL98} may be called `preonic'. Its
quantization was performed in \cite{BLS99}, where it was shown that the spectrum of the
quantized $n=4$ preonic superparticle is given by a tower of massless conformal higher spin
fields of all possible helicities and where evidence in favor that the $n=8$ and $n=16$
models describe conformal higher spin theories in $D=6$ and $D=10$ spacetimes was presented.

An elegant form of the bosonic and fermionic higher spin equations in
$\Sigma^{(10|0)}$ tensorial space was given and studied in \cite{V01s,V01c}.
The explicit form of the conformal higher spin equations in  $D=6,10$ tensorial spaces
was extracted from the their tensorial space version in  \cite{BBdAST05}.
The superfield form of the equations for the supermultiplets of $D=6, 10$ massless
conformal fields in the ${\cal N}=1$ tensorial superspaces $\Sigma^{(36|8)}$ and
$\Sigma^{(126|16)}$,
 \begin{equation}
 \label{x+th=10D}
D=10\,\quad\Sigma^{(126|16)}\; : \quad  Z^{\cal M } := (X^{\alpha\beta}\, , \, \theta^\alpha)
\; , \quad \cases{ \alpha,\beta =1,...,16 , \cr X^{\alpha\beta}= {1\over 16}
x^m\tilde{\sigma}{}_m^{\alpha\beta}+{1\over 2\cdot 5!}
y^{m_1...m_5}\tilde{\sigma}{}_{m_1...m_5}^{\alpha\beta} \; , \cr m=0,1,\ldots 9}
\end{equation}
was given in  \cite{BPST04}. In particular, the ${\cal N}=1$ supermultiplets of
$D$=4,6,10 conformal higher spin fields are described by real scalar superfields on the
corresponding $n$=4,8,16 $\Sigma^{({n(n+1)\over 2}|n)}$ superspaces,
\begin{equation}
\label{scalar} \Phi (X^{\alpha\beta} , \theta^\gamma)=b(X)+f_\alpha(X)\,\theta^\alpha+
\sum_{i=2}^{n}\phi_{\alpha_1\cdots\alpha_i}(X)\,\theta^{\alpha_1}\cdots\theta^{\alpha_i}\;,
\end{equation}
obeying the following simple superfield equation \cite{BPST04}
\begin{eqnarray}
\label{hsSEq=1N} D_{[\alpha}D_{\beta ]} \Phi (X, \theta ) = 0 \; .
\end{eqnarray}
Here,
\begin{eqnarray}
\label{DfN=1} D_\alpha= {\partial \over \partial \theta^\alpha} + i \theta^{\beta}
\partial_{\beta\alpha}\; , \qquad
D_{\alpha\beta}=\partial_{\alpha\beta}:={\partial \over \partial X^{\alpha\beta}}\; , \qquad
 \end{eqnarray}
are the covariant derivatives on the  rigid tensorial superspace
$\Sigma^{({n(n+1)\over 2}|n)}$ that satisfy
\begin{eqnarray}
\label{(DD)=2id} \{D_\alpha ,D_\beta\}=2i\partial_{\alpha\beta}
\end{eqnarray}
which, we note in passing, exhibits the central extension structure of the superalgebras
of tensorial superspaces \cite{BdAPV=30-32} (see further \cite{CAIPB:00}).

    The tensorial superspace higher spin equations with $\cal{N}$-extended supersymmetry,
however, have not been studied yet\footnote{See \cite{KuzenkoHS} as
well as \cite{GatesHS} and refs. therein for the description of supersymmetric free
higher spin equations in standard superspace and \cite{S+S=98,ESS=02}
for the superfield form of Vasiliev's interacting higher spin equations
\cite{Misha88-89,Misha92,Misha03s} in the usual simple and extended superspaces.}.
In this paper we fill this gap by presenting the free ${\cal N}$=2,
${\cal N}$=4 and ${\cal N}$=8 supersymmetric conformal higher
spin equations in ${\cal N}$-extended tensorial
superspaces $\Sigma^{({n(n+1)\over 2}|{\cal N}\,n)}$.
In particular, we show that the  ${\cal N}=2$
supermultiplets of $D=4,6$ and $10$ conformal higher spin equations are described by scalar,
chiral superfields $\Phi (X^{\alpha\beta} , \Theta^\gamma, \bar{\Theta}{}^\gamma)$ in ${\cal
N}=2$ tensorial superspace $\Sigma^{({n(n+1)\over 2}|2n)}$, which obey the following set of
linear superfield equations
\begin{eqnarray}
\label{hsSEq=2N} \bar{{\cal D}}_\alpha \Phi =0\; , \qquad {\cal D}_{[\alpha}{\cal D}_{\beta
]} \Phi  = 0 \; ,
\end{eqnarray}
where
\begin{eqnarray}
\label{Df-N=2} {\cal D}_\alpha &=& {\partial \over \partial \Theta^\alpha} + i
\bar{\Theta}^{\beta}
\partial_{\beta\alpha}= {1\over 2} (D_{\alpha 1}
+ i D_{\alpha 2})\, , \qquad \nonumber \\ \bar{{\cal D}}_\alpha &=& {\partial \over \partial
\bar{\Theta}^\alpha} + i {\Theta}^{\beta} \partial_{\beta\alpha}= - ({\cal D}_\alpha)^*\, ,
\qquad  {\cal D}_{\alpha\beta}=\partial_{\alpha\beta}:={\partial \over \partial
X^{\alpha\beta}}\; , \qquad
 \end{eqnarray}
are the covariant derivatives on the  the rigid $\mathcal{N}$=2
extended superspace $\Sigma^{({{n(n+1)}\over 2}|2n)}$, which obey the superalgebra
\begin{eqnarray}
\label{(DD)=2id} \{{\cal D}_\alpha , {\cal D}_\beta \}= 0\; , \qquad \{{\cal D}_\alpha
, \bar{{\cal D}}_\beta \}=2i\partial_{\alpha\beta} \; ,  \qquad \{\bar{{\cal D}}_\alpha ,
\bar{{\cal D}}_\beta \}=0 \; .
 \end{eqnarray}
The superfield equations in extended tensorial superspaces with even ${\cal N}> 2$
are given by a straightforward generalization of (\ref{hsSEq=2N})
(Eqs.~(\ref{*DPhi=0}), (\ref{asDDPhi=0}) in Sec.~3).

We also present the extended supersymmetric ${\cal N}>1$ generalization of the preonic
superparticle model of \cite{BLS99} and show how the above superfield equations can be
obtained by quantizing the $\Sigma^{({n(n+1)\over 2}|{\cal N}\,n)}$ superparticle model for
even ${\cal N}\geq 2$.

Although our superfield equations make sense for general  even ${\cal N}$ and $n$,
we elaborate in detail the ${\cal N}=2,4,8$ cases which, when $n=4$, correspond to the
supermultiplets of $D=4$ massless conformal higher spin theories with a clear
standard  `lower spin' field theory counterparts. These are  the hypermultiplet
for ${\cal N}=2$, the supersymmetric Yang-Mills supermultiplet for ${\cal N}=4$,
and the maximal supergravity multiplet for ${\cal N}=8$,
the linearized versions of which can be described by scalar superfields in the
standard  extended $D=4$ superspaces $\Sigma^{(4|4{\cal N})}$ with ${\cal N}=2,4,8$.

One of the reasons of our interest in $\mathcal{N}$-extended supersymmetric
systems of higher spin theories comes from the observation that $\mathcal{N}$-extended
supersymmetry with $\mathcal{N}=4$ unifies the scalar and vector gauge fields.
On the other hand, all the higher spin equations have been formulated in terms
of scalar and spinor fields in tensorial space, so that the study of
$\mathcal{N}$-extended supersymmetries might prove convenient in a search for a sensible
generalization of Maxwell and Einstein equations in tensorial superspace.
Indeed, at some point our study of the superfield equations in  ${\cal N}=4$
tensorial superspace $\Sigma^{(4|16)}$  produces a tensorial space counterpart
of the Maxwell equations. However, a careful analysis shows that the corresponding
(spin-)tensorial fields can be expressed as derivatives of other scalar fields in tensorial
superspace so that, for instance, the bosonic fields of ${\cal N}=4$ conformal higher
spin multiplet are actually given by two complex scalar fields in tensorial space,
$\phi$ and $\tilde{\phi}$. Similarly, when studying the superfield equations in
${\cal N}=8$  tensorial superspace $\Sigma^{(4|32)}$, although the tensorial
space generalizations of the conformal (super)gravity equations do appear
at an intermediate stage, it will turn out that they reduce to the scalar (and spinor)
field equations in the tensorial space form first presented by Vasiliev \cite{V01s}.
Roughly speaking one can state that the increasing of ${\cal N}$ results
just in multiplication of the scalar and spinor fields.

However, at ${\cal N}=4$ a new phenomenon does occur. As the new scalar field
appears in the theory only through a `Maxwell-like' field, which
is to say under the action of bosonic derivatives $\partial_{\alpha\beta}$, the
(free higher spin ${\cal N}=4$) theory becomes invariant under constant shifts
of this bosonic field which makes it similar to the axion (for which such a
symmetry is called Peccei-Queen \cite{Pe-Qui77} symmetry\footnote{As far as in type IIB string theory
and supergravity the axion appears as a member of the family of the RR gauge fields, its
Peccei-Queen symmetry can be considered as a counterpart of the gauge
symmetries characteristic of higher RR gauge potentials.}).  For the
${\cal N}=8$ multiplet  reformulated in terms of scalar and spinor fields,
the  Peccei-Queen symmetry becomes more complicated for the scalars and
is also present for spinor fields entering the model under the action of derivative
in the combination simulating the structure of the Rarita-Schwinger fields.

The plan of this paper is as follows. After discussing in Sec.~\ref{superPart}
the  $\Sigma^{({n(n+1)\over 2}|{\cal N}n)}$ superparticle model and the
structure of its constraints, we sketch its quantization  in Sec.~\ref{q-part}
(using the ${\cal N}=1$ results  of \cite{BLS99} to simplify the discussion)
and find our superfield equations, Eqs. (\ref{*DPhi=0}) and (\ref{asDDPhi=0}),
as the condition obeyed by the superparticle wavefunction
in the tensorial superspace coordinate representation.
In Sec.~4 we study the field content of our superfield equations, which
are valid for arbitrary even  ${\cal N}$ and $n$ (with $n=4,8,16$ corresponding
to $D=4,6,10$ free massless conformal higher spin theories) for
the particular cases of ${\cal N}= 2,4,8$. In particular, we show in
Sec.~\ref{Nigual2} (where the analysis is not restricted to $n=4$)
that the  ${\cal N}=2$ supermultiplets $D=4,6$ and $10$ of conformal higher spin fields
described by complex scalar and complex spinor ($s$-vector) fields
obeying the Vasiliev's tensorial space higher spin equations are
encoded in the chiral scalar superfield
$\Phi (X^{\alpha\beta} , \Theta^\gamma, \bar{\Theta}{}^\gamma)$ on
${\cal N}=2$ tensorial superspace $\Sigma^{({n(n+1)\over 2}|2n)}$ which obeys the
set of linear superfield equations (\ref{hsSEq=2N}). In Sec.~(\ref{secN4})
we study the superfield equations (\ref{*DPhi=0}) and (\ref{asDDPhi=0}) in
${\cal N}=4$ tensorial superspace, find a tensorial superspace counterpart of
$D=4$ Maxwell equation and show that its general solution is expressed through the
scalar field obeying the tensorial space higher-spin equation of usual type.
We also show there that this new scalar field is defined up to a constant shift
which resembles the Peccei-Quinn transformation of the axion. Sec.~\ref{N8part} analyzes
the field content of the complex superfield obeying a ${\cal N}=8$  version of
Eqs.~(\ref{*DPhi=0}), (\ref{asDDPhi=0}) and presents the more complicated
Peccei-Quinn-like symmetry characteristic of the ${\cal N}=8$ massless
conformal free higher spin theories. Some comments on possible extensions
of this work are made in Sec.~\ref{conclu}.

\bigskip

\section{Superparticle in $\mathcal{N}$-extended tensorial superspace}
\label{superPart} \setcounter{equation}0

\subsection{An action for the $\Sigma^{({n(n+1)\over 2}|{\cal N}\,n)}$ superparticle}

The superparticle action in $\Sigma^{({n(n+1)\over 2}|{\cal N}\,n)}$ has the form
\begin{eqnarray}
\label{SNpreon} S =  \int d \tau \, {\cal L}  =  \int d \tau \,
[\dot{\hat{X}}{}^{\alpha\beta}(\tau) - i \dot{\hat{\theta}}{}^{\alpha I}(\tau)
\hat{\theta}{}^{\beta I}(\tau)]\lambda_{\alpha}(\tau)\lambda_{\beta}(\tau)\; ,  \qquad
\cases{\alpha , \beta = 1,...,n \; , \cr I=1,2,..., {\cal N}\; ,}
\end{eqnarray}
where the $\lambda_\alpha(\tau)$ are auxiliary {\it commuting} spinor variables,
$\hat{X}^{\alpha\beta}(\tau)=\hat{X}^{\beta\alpha}(\tau)$ and $\hat{\theta}^{\alpha I}(\tau)$
are the bosonic and fermionic coordinate functions that define the superparticle worldline
$W^1\in \Sigma^{({n(n+1)\over 2}|{\cal N}\,n)}$,
$\hat{Z}^{\cal M}(\tau)=(\hat{X}{}^{\alpha\beta}(\tau)\,,\,\hat{\theta}{}^{\alpha I}(\tau))$,
and the dot denotes derivative with respect to proper time $\tau$;
due to the time derivative, the term
$\dot{\hat{\theta}}{}^{\alpha I}(\tau) \hat{\theta}{}^{\beta I}(\tau)$
is also ($\alpha \leftrightarrow \beta$)-symmetric.

It is convenient to write the action (\ref{SNpreon}) in the form
\begin{eqnarray}
\label{SNpr=} S =\int_{W^1} {\hat{\Pi}}{}^{\alpha\beta}
\lambda_{\alpha}(\tau)\lambda_{\beta}(\tau)\; ,  \qquad
\end{eqnarray}
where
\begin{eqnarray}
\label{hPi=}
{\hat{\Pi}}{}^{\alpha\beta}(\tau) = d\tau {\hat{\Pi}}{}_\tau^{\alpha\beta}(\tau)
= d\tau( {\dot{\hat{X}}{}^{\alpha\beta} - i \dot{\hat{\theta}}{}^{I (\alpha} }
\hat{\theta}{}^{\beta ) I} )\; , \quad  \cases{\alpha , \beta = 1,...,n \; ,
\cr I=1,2,..., {\cal N}\; ,}
\end{eqnarray}
is the pull-back to $W^1$ of the vielbein $\Pi{}^{\alpha\beta}$
of the flat ${\cal N}$-extended tensorial superspace
$\Sigma^{({n(n+1)\over 2}|{\cal N}n)}\,$ (the bosonic Maurer-Cartan
one-form on the $\Sigma^{({n(n+1)\over 2}|{\cal N}n)}\,$
supergroup manifold), namely
\begin{eqnarray}
\label{Pi=} {{\Pi}}{}^{\alpha\beta}= d{{X}}{}^{\alpha\beta} -id{{\theta}}{}^{I (\alpha }
{\theta}{}^{\beta )I}\; ,  \quad
\end{eqnarray}
by the map $\phi : W^1 \rightarrow \Sigma^{({n(n+1)\over 2}|{\cal N}\,n)}\,$
($\phi^*(\Pi{}^{\alpha\beta})\equiv \hat{\Pi}{}^{\alpha\beta}(\tau)$).

The superparticle action is manifestly invariant under the supertranslations on the rigid
${\cal N}$-extended tensorial superspace $\Sigma^{({n(n+1)\over 2}|{\cal N}n)}\,$,
\begin{eqnarray}
\label{trans=}  \delta_a {{X}}{}^{\alpha\beta} =
a{}^{\alpha\beta}\quad , \quad  \delta_a {{\theta}}{}^{I \alpha }=0\; , \quad
\\ \label{susy=} \delta_\epsilon
{{X}}{}^{\alpha\beta} = i{{\theta}}{}^{I (\alpha }{\epsilon}{}^{\beta )I}\quad , \quad
\delta_\epsilon {{\theta}}{}^{I \alpha } = {\epsilon}{}^{\beta I}\quad ,
\end{eqnarray}
which act on the worldline fields as
\begin{eqnarray}
\label{trans=h}
\delta_a {\hat{X}}{}^{\alpha\beta} (\tau) =
a{}^{\alpha\beta}\;,\quad \delta_a {\hat{\theta}}{}^{I \alpha }= 0\; ;
 \qquad \delta_a \hat{\lambda}_\alpha =0 \; . \\
\label{susy=h}
 \delta_\epsilon {\hat{X}}{}^{\alpha\beta}
 =i{\hat{\theta}}{}^{I (\alpha }{\epsilon}{}^{\beta )I}\; , \quad
\delta_\epsilon {\hat{\theta}}{}^{I \alpha }= {\epsilon}{}^{\beta I}\; ; \quad
\delta_\epsilon {\lambda}_\alpha =0 \; .
\end{eqnarray}
The action (\ref{SNpreon}) is also manifestly invariant  under the $GL(n, \bb{R})$
transformations of the $\alpha, \beta=1,...,n$ indices, which reduce to the
$n$-dimensional representation of $Spin(1,D-1)$ when these indices are thought
of as Lorentz-spinorial ones\footnote{In \cite{V01s,V01c} the counterparts of
${\lambda}_\alpha$ were called `$s$-vectors' to avoid their immediate
identification as $GL(n,\bb{R})$ vectors or $SO(1,D-1)$ spinors. }.

\subsection{Symplectic supertwistor form of the action}

Actually, the  $\Sigma^{({n(n+1)\over 2}|{\cal N}n)}\,$ superparticle
action (\ref{SNpreon}) is invariant under the larger $OSp({\cal N}|2n)$ supergroup.
To make this manifest as well as to determine easily the number
of physical degrees of freedom it is convenient to use Leibniz's rule\footnote{It is
sufficient to use $\lambda_{\alpha}\lambda_{\beta}d{{X}}{}^{\alpha\beta}=
\lambda_{\alpha}d(\lambda_{\beta}{{X}}{}^{\alpha\beta}) - \lambda_{\alpha}
{{X}}{}^{\alpha\beta}d\lambda_{\beta}$; no integration by parts is needed.} to rewrite the
action (\ref{SNpreon}) in the form
\begin{eqnarray}
\label{SNpr=STw} S = \int_{W^1}   (\lambda_{\alpha}d\mu^\alpha - \mu^\alpha d\lambda_{\alpha}
- id\chi^I\, \chi^I ) = \int_{W^1} d\Upsilon^{\Sigma}\Xi_{\Sigma\Omega}\Upsilon^{\Omega}\; .
\qquad
\end{eqnarray}
This action is written in terms of the bosonic spinor $\lambda_\alpha(\tau)$,
which is present in (\ref{SNpreon}),  a second bosonic spinor $\mu^\alpha$ and $\mathcal{N}$ real
fermionic variables $\chi^I$; these form the $\mathcal{N}$-extended, ($2n+{\cal N}$)-dimensional
orthosymplectic supertwistor (see \cite{BL98,BLS99} and also \cite{BdAPV=30-32} for ${\cal N}=1$)
\begin{eqnarray}
\label{YS=}
\Upsilon^{\Sigma}= \left(\matrix{\mu^\alpha \cr \lambda_{\alpha}\cr   \chi^I
}\right)\; , \qquad \alpha =1,\ldots, n \; , \qquad I=1,\ldots, {\cal N} \; .
\end{eqnarray}
This generalizes the Penrose twistors \cite{Penrose} (or conformal $SU(2,2)$ spinors)
and the Ferber-Schirafuji supertwistors \cite{Ferber,Shirafuji} (carrying the basic
representation of $D$=4 $SU(\mathcal{N}|2,2)$). The $\Upsilon^{\Sigma}$'s carry the defining
representation of the $OSp({\cal N}|2n)$ supergroup (see in this
context \cite{BL98, BLS99,BdAPV=30-32} for ${\cal N}=1$),
the transformations of which preserve the $(2n+\mathcal{N})\times (2n+\mathcal{N})$
orthosymplectic `metric' $\Xi_{\Sigma\Omega}$,
\begin{eqnarray}
\label{YS=} \Xi_{\Sigma\Omega}= \left(\matrix{0 &   \delta_\alpha{}^\beta & 0 \cr -
\delta^{\alpha}{}_\beta & 0 & 0 \cr  0 & 0 & -i\delta^{IJ} }\right)\; , \quad \alpha
=1,\ldots, n \; , \quad I=1,\ldots, {\cal N} \; .
\end{eqnarray}
In fact, $OSp(1|2n)$ may be
considered as a supersymmetric generalization of the superconformal group for $D={n\over 2}+2$
(see \cite{vH+vP=82,BL98,BLSP2000,BdAPV=30-32} and refs. therein).

The relations between the supertwistor components and the variables of the action
(\ref{SNpreon}) that make the transition between both actions are
\begin{eqnarray}
\label{mu=} \mu^\alpha = {\hat{X}}{}^{\alpha\beta} \lambda_{\beta} - {i\over 2}
{\hat{\theta}}{}^{\alpha I} \chi^I  \; , \qquad \chi^I=  {\hat{\theta}}{}^{\alpha I}\,
\lambda_\alpha \; , \qquad
\end{eqnarray}
which generalize the Penrose and the Ferber-Shirafuji incidence relations
\cite{Penrose,Ferber,Shirafuji} (see \cite{BL98} and
\cite{BLS99,BdAPV=30-32} for ${\cal N}=1$). Since the action (\ref{SNpr=STw})
does not possess any gauge symmetries, the components of the orthosymplectic
supertwistors are the true, physical degrees of freedom ($2n$ bosonic and ${\cal N}$
fermionic) of our generalized superparticle model.

\medskip

\subsection{Gauge symmetries}

By construction, the actions (\ref{SNpr=STw}) and (\ref{SNpreon})  describe the same
dynamical system. Thus, since the action (\ref{SNpreon}) depends on ${n(n+1)\over 2}+n$
bosonic variables and $\mathcal{N}n$ fermionic ones, it should possess $n(n-1)/2$
bosonic gauge symmetries and ${\cal N}\, (n-1)$ fermionic ones to reduce the
number of degrees of freedom to those of the supertwistors $\Upsilon^{\Sigma}$
appearing in the action (\ref{SNpr=STw}). The simplest way to describe these gauge symmetries,
called  fermionic $\kappa$-symmetry and bosonic $b$-symmetry,  is to define
restrictions on the basic variations of the bosonic and fermionic coordinate
functions (see \cite{BLS99,BdAPV=30-32} for the ${\cal N}=1$ case)
\begin{eqnarray}
\label{k-sym=a} \delta_\kappa  {\hat{X}}{}^{\alpha\beta}= i \delta_\kappa
{\hat{\theta}}{}^{I(\alpha } {\hat{\theta}}{}^{\beta) I}\; , \qquad \fbox{$\delta_\kappa
{\hat{\theta}}{}^{\alpha I}\lambda_\alpha =0 $}\; , \qquad \delta_\kappa \lambda_\alpha =0 \;
, \qquad \\ \label{k-sym=b} \fbox{$ \delta_b  {\hat{X}}{}^{\alpha\beta} \lambda_\alpha =0
$}\; , \qquad \delta_b {\hat{\theta}}{}^{I\alpha }\lambda_\alpha =0\; , \qquad \delta_b
\lambda_\alpha =0 \; . \qquad
\end{eqnarray}

\medskip

\subsection{Constraints and their conversion to first class}

In the hamiltonian formalism, the $\delta_\kappa$ and $\delta_b$ gauge symmetries in
Eqs.~(\ref{k-sym=a}), (\ref{k-sym=b}) are generated by first class constraints which may be
extracted from the following bosonic and fermionic primary constraints of the model
(\ref{SNpreon})
\begin{eqnarray}
 \label{Df=pi-}
 {d}_{\alpha I} := \pi_{\alpha I} + i P_{\alpha\beta}  \theta^{\beta I} \approx 0 \; ,
 \qquad \;  \\
 \label{P-ll=} {} {\cal P}_{\alpha\beta} := P_{\alpha\beta} - \lambda_\alpha\lambda_\beta
\approx 0\; , \qquad P^{\alpha (\lambda )} \approx 0\; , \qquad
\end{eqnarray}
where
\begin{eqnarray}
\label{P:=} P_{\alpha\beta}:= {1\over 2}{\delta {\cal L} \over \delta
\dot{\hat{X}}{}^{\alpha\beta}} \; , \qquad
P_\alpha^{(\lambda )}:= {\delta {\cal L} \over \delta \lambda{}^{\alpha}}\; , \qquad
\pi_{\alpha I}:= {\delta {\cal L} \over \delta \dot{\hat{\theta}}{}^{\alpha}} \; ,  \qquad
\end{eqnarray}
are the canonical momenta conjugated to the coordinate functions and to the auxiliary bosonic
spinor ($s$-vector).  Using the canonical Poisson brackets
\begin{eqnarray}
 \label{Poisson=b}
{} &&  [\hat{X}{}^{\gamma\delta} ,  P_{\alpha\beta}]_{PB} = - [P_{\alpha\beta} ,
\hat{X}{}^{\gamma\delta}]_{PB} = \delta_{\alpha }{}^{(\gamma}\delta_{\beta}{}^{\delta )} \, ,
\;\; {}[ \lambda{}_{\beta} ,  P^{\alpha (\lambda )}]_{PB}  =
  - [P^{\alpha (\lambda )} , \lambda{}_{\beta}]_{PB} =
\delta_{\beta}{}^{\alpha }
  \, ,  \qquad  \\
\label{Poisson=f}
&&  \{ \pi_{\alpha} \, , \, \hat{\theta}{}^{\beta}\}_{PB}=
 \{ \hat{\theta}{}^{\beta}\, , \, \pi_{\alpha} \}_{PB}= - \delta_{\alpha }{}^{\beta} \; ,
\end{eqnarray}
it follows that the nonvanishing Poisson brackets of the above constraints are
\begin{eqnarray}
\label{[C,C]=} {}\{ {d}_{\alpha I} , {d}_{\beta J} \}_{PB}= - 2i P_{\alpha\beta}  \delta_{IJ}
\; , \qquad [{\cal P}_{\alpha\beta}, P^{\gamma (\lambda)}]_{PB} =  -2\lambda_{(\alpha
}\delta_{\beta )} {}^\gamma
  \; . \qquad
\end{eqnarray}
These clearly indicate that the primary constraints above are a mixture of first and second
class constraints. Rather than separating them, we use below the so-called `conversion
procedure' (see \cite{BLS99} for references), by which a pair of degrees of freedom is added
to each pair of second class constraints to modify Eqs.~(\ref{[C,C]=}) in such a way that
they form a closed algebra. In this way, these modified constraints become first class ones
generating gauge symmetries in the enlarged phase space. In it, all the constraints of the
model are first class and account as well for the original second class constraints. These
can be recovered by gauge fixing the additional gauge symmetries/first class constraints of
the system in the enlarged phase space. For the ${\cal N}=1$ version of (2.1)
this was done in \cite{BLS99}.

As the bosonic sector of all the superparticle models is the same irrespective of ${\cal N}$,
we may use the results of \cite{BLS99} for ${\cal N}=1$ and state
that the conversion in the bosonic sector is effectively reduced to ignoring the
constraints $P^{\alpha (\lambda )}\approx 0$ in the analysis. An easy way
to see that this is indeed consistent is to observe that, as far as
$\lambda_\alpha\not= (0,...,0)$ (the usual configuration space restriction for twistor-like
variables), the second brackets in (\ref{[C,C]=}) show that $ P^{\alpha (\lambda )}= 0$ is a
good gauge fixing condition for $n$ of $n(n+1)/2$ gauge symmetries generated by the
constraints ${\cal P}_{\alpha\beta}$.

To perform the conversion in the fermionic sector, we introduce the ${\cal N}$ fermionic
variables $\chi^I$ and postulate for them the Clifford-type Poisson brackets
\begin{eqnarray}
 \label{chi,chi=}
{}\{ {\chi}^{I} , {\chi}^{J} \}_{PB}= - 2i   \delta^{IJ}  \; . \qquad
\end{eqnarray}
These ${\chi}^{I}$ are then used to modify the fermionic constraints to ${\bb D}_{\alpha I}
=d_{\alpha I} + i\chi^I\lambda_\alpha$. Thus, after conversion, the superparticle model
(\ref{SNpreon}) is described by the following set of first class constraints
\begin{eqnarray}
 \label{convC=}
{\bb D}_{\alpha I}  := \pi_{\alpha I} + i P_{\alpha\beta}  \theta^{\beta I} + i\chi^I
\lambda_\alpha \approx 0\; , \qquad {\cal P}_{\alpha\beta} := P_{\alpha\beta} -
\lambda_\alpha\lambda_\beta \approx 0\; , \qquad
\end{eqnarray}
which obey the superalgebra of the rigid supersymmetry of $\mathcal{N}$-extended tensorial superspace
$\Sigma^{({n(n+1)\over 2}|{\cal N}n)}$ in Eq.~(\ref{(DD)=2id}), namely
\begin{eqnarray}
 \label{[C,C]=1}
{}\{ {\bb D}_{\alpha I} , {\bb D}_{\beta J} \}_{PB}= -2i {\cal P}_{\alpha\beta}  \delta_{IJ}
\; , \qquad [ {\cal P}_{\alpha\beta}, {\bb D}_{\gamma I} ]_{PB}= 0 \; , \qquad [ {\cal
P}_{\alpha\beta}, {\cal P}_{\gamma\delta}]_{PB} =  0
  \; . \qquad
\end{eqnarray}

\medskip

\section{Quantization of the superparticle in $\Sigma^{({n(n+1)\over 2}|\mathcal{N}n)}$ with even
$\mathcal{N}$ and conformal higher spin equations} \label{q-part}
\setcounter{equation}0

Quantizing the model in its orthosymplectic-twistorial formulation (\ref{SNpr=STw}) is
straightforward. The canonical hamiltonian is equal to zero and thus the Schr\"odinger
equation simply states that
the wavefunction is independent of $\tau$. Following a procedure similar to that in
\cite{BLS99} one can show that, in the $n=4$ tensorial space corresponding to  $D=4$, the
wavefunction of the bosonic limit of the superparticle model (\ref{SNpr=STw}) describes the
solution of the free higher spin equations. This means that it can be written in terms of an
infinite tower of left and right chiral fields $\phi_{A_1\ldots A_{2s}}(p_\mu)$ and
$\phi_{\dot{A}_1\ldots \dot{A}_{2s}}(p_\mu)$ for all half-integer values of $s$ with $p_\mu
p^\mu=0$.

Let $\mathcal{N}>1$ and even (as it will be henceforth). Quantization {\it \`a la} Dirac of a
dynamical system with first class constraints requires imposing them as equations to be
satisfied by its wavefunction. In the case of our superparticle model (\ref{SNpreon}) this
wavefunction can be chosen to depend on the coordinates of $\mathcal{N}$-extended tensorial
superspace ($X^{\alpha\beta}=X^{\beta\alpha}$, $\theta^{\alpha I}$), on the bosonic spinor
($s$-vector) variable $\lambda_\alpha$ and on {\it a half} of the fermionic variables
$\chi^I$ as far as they are, by (\ref{chi,chi=}), their own momenta. The separation of a half
of the real $\chi^I$ coordinates can be achieved by introducing complex\footnote{A separation
in pairs of conjugate constraints is used in the Gupta-Bleuler
method of quantizing systems with second class constraints as the massive
superparticle \cite{dAz-Luk:82,Fryd:84}.} variables $\eta^q$,
$(\eta^q)^*=\bar{\eta}_q$, $q=1,\ldots , {\cal N}/2$, so that $\chi^I=(\chi^q , \chi^{{\cal
N}/2 +q})= ((\eta^q+\bar{\eta}_q) , i(\bar{\eta}_q-{\eta}^q))$,
\begin{eqnarray}
\label{etai=} \eta^q= {{\chi}^q -  i\chi^{{\cal N}/2 +q}\over 2}\; , \qquad \bar{\eta}_q =
{\chi^q + i\chi^{{\cal N}/2 +q}\over 2}\; , \qquad \{ \bar{\eta}_q,
\eta^p\}_{PB}=- i\delta_q{}^p \; . \qquad
\end{eqnarray}
Then, the wavefunction superfield in the coordinates representation depends only on
$\eta^q$,
\begin{eqnarray}
 \label{cW=}
 {\cal W}= {\cal W}(X^{\alpha\beta},\theta^\alpha ; \lambda_\alpha ; \eta^q)\; ,
\end{eqnarray}
the various momenta are given by the differential operators
\begin{eqnarray}
 \label{opP=}
 P_{\alpha\beta}=- i \partial_{\alpha\beta} \; ,\qquad
 \pi_{\alpha I}=-i {\partial\over \partial {\theta}^{\alpha I}} \;,
 \qquad \bar{\eta}_q =  {\partial \over \partial \eta^q} \; ,
\end{eqnarray}
and the Poisson brackets become quantum commutators or anticommutators ($[\;,\;\}_{PB}\mapsto
\frac{1}{i\hbar}[\;,\;\}$; we take $\hbar= 1$). The quantum constraints operators, to be
denoted by the same symbol (although having in mind ${\bb D}_{\alpha I}\mapsto -i {\bb
D}_{\alpha I}^{quantum}$, ${\cal P}_{\alpha\beta}^{classical}\mapsto -i {\cal
P}_{\alpha\beta}^{quantum}$) are then
\begin{eqnarray}
\label{bbDI=} {\bb D}_{\alpha I}  := {\partial\over \partial {\theta}^{\alpha I}} + i
\partial_{\alpha\beta}  \theta^{\beta I} - \chi^I \lambda_\alpha \; , \qquad \\ \label{cP=qu}
{\cal P}_{\alpha\beta} := \partial_{\alpha\beta} -i \lambda_\alpha\lambda_\beta\; , \qquad
\end{eqnarray}
and have to be imposed on the wavefunction (\ref{cW=}).

For even $\mathcal{N}$, it is convenient to introduce complex
Grassmann coordinates and complex Grassmann derivatives,
\begin{eqnarray}
\label{qTh=}
\Theta^{\alpha q} = {1\over 2}(\theta^{\alpha q}-i \theta^{\alpha (q+{\cal
N}/2)})= (\bar{\Theta}^{\alpha}_q)^*  \quad \Leftrightarrow \quad \partial_{\alpha q}:=
{\partial\over \partial {\Theta}^{\alpha q}}={\partial\over \partial {\theta}^{\alpha q}} + i
{\partial\over \partial {{\theta}}^{\alpha (q+{\cal N}/2)}}\; ,
\end{eqnarray}
$q=1,\dots,\mathcal{N}/2$, and conjugate pairs of fermionic constraints
\begin{eqnarray}
 \label{nablaq=}
 {\nabla}_{\alpha q} & := & {\bb D}_{\alpha q}+ i {\bb D}_{\alpha (q+{\cal N}/2)}  =
 \partial_{\alpha q} + 2i \partial_{\alpha\beta}  \bar{\Theta}{}^{\beta}_{q} -
 2  \lambda_\alpha \, {\partial \over \partial \eta^q}=: {\cal D}_{\alpha q} -
 2  \lambda_\alpha \, {\partial \over \partial \eta^q}  \; , \qquad \\
\label{bnablaq=} \bar{\nabla}_{\alpha}{}^{q} & := & {\bb D}_{\alpha q}- i {\bb D}_{\alpha
(q+{\cal N}/2)}  = \bar{\partial}_{\alpha}{}^{q} + 2i \partial_{\alpha\beta}
{\Theta}{}^{\beta q} - 2  \lambda_\alpha \, \eta^q=: {\bar{\cal D}}{}_{\alpha}{}^{ q} -2
\lambda_\alpha  \eta^q  \; . \qquad
\end{eqnarray}
Since $\{ {\cal D}_{\alpha q} ,\bar{{\cal D}}_{\beta}^{p} \}= 4i {\partial}_{\alpha\beta}
\delta_{q}^{p}\,$, the above ${\nabla}_{\alpha q}$, $\bar{\nabla}_{\alpha}{}^{q}$ and the
bosonic constraint (\ref{cP=qu}) determine the superalgebra given by the only nonzero
bracket
\begin{eqnarray}
 \label{q[C,C]=1}
\{ {\nabla}_{\alpha q} ,\bar{{\nabla}}_{\beta}^{p} \}= 4i {\cal P}_{\alpha\beta}
\delta_{q}^{p} \quad.
\end{eqnarray}
This shows that it is sufficient to impose on the superwavefunction (\ref{cW=}) the fermionic
constraints,
\begin{eqnarray}
 \label{CfW=0}
{\nabla}_{\alpha q}  {\cal W}:=  {\cal D}_{\alpha q}  {\cal W} - 2  \lambda_\alpha \,
{\partial \over \partial \eta^q}  {\cal W}=0 \, , \qquad \\
 \label{Cf*W=0}
\bar{{\nabla}}_{\alpha}^{p} {\cal W}:=\bar{\cal D}{}_{\alpha}^{ q}{\cal W} - 2  \lambda_\alpha
\, \eta^q {\cal W}=0 \; , \qquad
\end{eqnarray}
since the mass-shell-like bosonic constraint,
\begin{eqnarray}
 \label{CbW=0}
{\cal P}_{\alpha\beta}   {\cal W}:= (\partial_{\alpha\beta} - i \lambda_\alpha\lambda_\beta
)   {\cal W}=0
  \; , \qquad
\end{eqnarray}
will follow as a consistency condition for (\ref{CfW=0}), (\ref{Cf*W=0}).

Decomposing the superwavefunction in a finite power series in the complex Grassmann variable
$\eta^q$,
\begin{eqnarray}
 \label{cW=W0+etaqWq+}
  {\cal W} (X,\Theta^q,\bar{\Theta}_q,\lambda, \eta^q) =
 W^{(0)}(X,\Theta^q,\bar{\Theta}_q,\lambda) +
 \sum\limits_{k=1}^{{\cal N}/2} {1\over k!} \eta^{q_k}\ldots \eta^{q_1} \,
 W^{(k)}_{q_1\ldots q_k} (X,\Theta^q,\bar{\Theta}_q,\lambda)
  \; , \qquad
\end{eqnarray}
we find that Eqs.~(\ref{CfW=0}), (\ref{Cf*W=0}) imply
\begin{eqnarray}
 \label{DW=lW}
 {\cal D}_{\alpha q} W^{(0)}= 2\lambda_\alpha W_q^{(1)}\; , \quad ... \; , \quad {\cal D}_{\alpha q}
 W_{q_1\ldots q_k}^{(k)}=
 2\lambda_\alpha  W_{qq_1\ldots q_k}^{(k+1)}\;  , \quad \ldots \; , \qquad \nonumber \\
 {\cal D}_{\alpha q} W_{q_1\ldots q_{{\cal N}/2}}^{({\cal N}/2)}=  0\; ,
\end{eqnarray}
and
\begin{eqnarray}
 \label{*DW=lW}
 \bar{{\cal D}}{}_{\alpha}^{q} W^{(0)}= 0\; , \quad
 \bar{{\cal D}}{}_{\alpha}^{q}  W_{q_1\ldots q_k}^{(k)}=
 2k \lambda_\alpha  W_{[q_1\ldots q_{k-1}}^{(k-1)}\delta_{q_k]}{}^q\; , \quad \ldots \; ,
 \quad \nonumber \\
 \bar{{\cal D}}{}_{\alpha}^{q} W_{q_1\ldots q_{{\cal N}/2}}^{({\cal N}/2)} =
 {\cal N}  \lambda_\alpha  W_{[q_1\ldots q_{{\cal N}/2-1}}^{({\cal N}/2-1)}\delta_{q_{{\cal
 N}/2}]}^q
  \; .
\end{eqnarray}
Eqs.~(\ref{DW=lW}) show that all the superfields $W_{q_1\ldots q_k}^{(k)}$ can be constructed
from fermionic derivatives of the superfield $W^{(0)} (X,\Theta^q,\bar{\Theta}_q,\lambda)$
which is chiral as a consequence of the first equation in  (\ref{*DW=lW}),
\begin{eqnarray}
\label{DDDW=lllW}
 {\cal D}_{\alpha_{k} q_{k}}\ldots  {\cal D}_{\alpha_1 q_1} W^{(0)}=
 2^k \lambda_{\alpha_1} \ldots \lambda_{\alpha_k} W_{q_1\ldots q_k}^{(k)}
  \; , \qquad  \bar{{\cal D}}{}_{\alpha}^{q} W^{(0)}= 0  \; . \qquad
\end{eqnarray}
Then, the  wavefunction ${\cal W}$ is completely characterized by the chiral superfield
$W^{(0)} (X,\Theta^q,\bar{\Theta}_q,\lambda)$. As a consequence of (\ref{CbW=0}), $W^{(0)}$
obeys
\begin{eqnarray}
 \label{CbW(0)=0}
{\cal P}_{\alpha\beta}  W^{(0)}:= (\partial_{\alpha\beta} - i \lambda_\alpha\lambda_\beta  )
W^{(0)} =0 \; ,
\end{eqnarray}
which is solved by a planewave in tensorial space,
\begin{eqnarray}
 \label{W0=eXll}
 W^{(0)} (X,\Theta^q,\bar{\Theta}_q,\lambda) =
 \tilde{W} (\lambda, \Theta^q,\bar{\Theta}_q)\exp \{ i \lambda_\alpha\lambda_\beta
 X^{\alpha\beta}\}\; .
\end{eqnarray}
Then, the chirality of $W^{(0)}$ and the first equation in  (\ref{DW=lW}), which now implies
$\partial_{\alpha q} W^{(0)}\propto \lambda_\alpha$, show that the general solution for the
superparticle wavefunction is determined by the following chiral plane wave superfield
\begin{eqnarray}
 \label{W0=WeXll}
 W^{(0)} (X,\Theta^q,\bar{\Theta}_q,\lambda) = {w}(\lambda \, , \, \Theta^q\lambda)
 \exp \{ i \lambda_\alpha\lambda_\beta (X+2i\Theta^{p}\bar{\Theta}_p)^{\alpha\beta}\}\; ,
 \qquad
\end{eqnarray}
where $\Theta^q\lambda=\Theta^{\alpha\, q}\lambda_\alpha$ and
\begin{eqnarray}
 \label{w=}
 {w}(\lambda \, , \, \Theta^q\lambda )  = w^{(0)}(\lambda ) +
 \sum\limits_{k=1}^{{\cal N}/2} {1\over k!} (\lambda\Theta^{q_k})\ldots (\lambda\Theta^{q_1})
 \,
 w^{(k)}_{q_1\ldots q_k} (\lambda)\; .  \qquad
\end{eqnarray}

\medskip

We refer to \cite{BLS99} for a discussion on how the arbitrary function
$w^{(0)}(\lambda_\alpha )$ with $\alpha=1,2,3,4$ encodes all the solutions
of the massless higher spin equations in $D=4$ and to \cite{BLS99,BBdAST05}
for the $D=6, 10$ cases. The key point is that $\lambda_\alpha$
carries the degrees of freedom of a light-like momentum
($\lambda {\gamma}_a\lambda$ is light-like in $D$=4,6,10 which corresponds
to $n$=4,8,16) plus those of spin. The $d.o.f.$ of $\lambda_\alpha$ and
those of the lightlike momenta are encoded, both up to a scale factor,
in the coordinates of the compact manifolds $S^{n-1}=S^{2D-5}$
and $S^{\frac{n}{2}}=S^{D-2}$, respectively. The spheres\footnote{The `celestial
spheres' $S^{D-2}$ are the bases $S^{2,4,8}$ of the Hopf fibrations
$S^{2D-5}\rightarrow S^{D-2}\; (S^{n-1}\rightarrow S^{n \over 2})\,$ of $S^{\,3,7,15}$,
($n,D$)=(4,4), (8,6), (16,10). The fibres $S^{D-3}=S^{1,3,7}$ of these bundles
correspond to the complex, quaternion and octonion numbers of unit modulus. The
remaining $n$=2, $D$=3 case corresponds to the first of the four Hopf fibrations,
$S^1\rightarrow \mathbb{R}P^1$; its fibre is determined by the reals of unit
modulus, $Z_2$, and there are no extra coordinates.} $S^{\frac{n}{2}-1}$
are related to helicity in the $n=4,\,D=4$ case and to its multidimensional
generalizations for $D=6,10$ \cite{BLS99,BBdAST05}.

To obtain a superfield on tensorial superspace describing massless conformal higher spin
theories with extended supersymmetry, the wavefunction  (\ref{W0=WeXll}) has to
be integrated over $\,{\bb R}^n- \{0\} \sim {\bb S}^{n-1}\times$${\bb R}_+ \,$,
parametrized by $\lambda_\alpha$, with an appropriate
measure that we denote by $d^{n}\lambda$,
\begin{eqnarray}
\label{Phi=intWeXll}
  \Phi (X,\Theta^q,\bar{\Theta}_q) =
  \int d^{n}\lambda W^{(0)} (X,\Theta^q,\bar{\Theta}_q,\lambda) =
   \int d^{n}\lambda {w}(\lambda \, , \, \Theta^q\lambda)
   e^{ i \lambda_\alpha\lambda_\beta (X+2i\Theta^{p}\bar{\Theta}_p)^{\alpha\beta}}\; .
\end{eqnarray}
One can easily check that the superfield $\Phi$ is chiral
\begin{eqnarray}
 \label{*DPhi=0}
  {\bar {\cal D}}_\alpha^q\Phi (X,\Theta^{q'},\bar{\Theta}_{p'}) = 0\;   \qquad
\end{eqnarray}
and satisfies the equation
\begin{eqnarray}
 \label{asDDPhi=0}
  {\cal D}_{q[\beta}{\cal D}_{\gamma ]p}\Phi (X,\Theta^{q'},\bar{\Theta}_{p'}) = 0\;  .
  \qquad
\end{eqnarray}
These are the superfield equations for the wavefunction of the superparticle in ${\cal
N}$-extended tensorial superspace for even ${\cal N}$.

\section{From the superfield to the component form of the higher spin equations
in tensorial space} \label{sec4}
\setcounter{equation}0

\subsection{${\cal N}=2$}
\label{Nigual2}

When ${\cal N}=2$, Eqs.~(\ref{*DPhi=0}) and (\ref{asDDPhi=0}) coincide with
Eqs.~(\ref{hsSEq=2N}). It is easy to check that Eq.~(\ref{asDDPhi=0}) with ${\cal N}=2$
implies the vanishing of all the components of the `chiral' superfield $\Phi (X,\Theta
,\bar{\Theta})= \Phi (X+2i\Theta \cdot \bar{\Theta} ,\Theta )$, except the first two,
\begin{eqnarray}
 \label{Phi=N2}
  \Phi (X,\Theta ,\bar{\Theta})= \phi(X_L )+ i\Theta^\alpha  \psi_\alpha (X_L)\quad ,  \quad
  X_L^{\alpha\beta}=X^{\alpha\beta}+2i\Theta^{(\alpha}\bar{\Theta}{}^{\beta )}
  =X_L^{\beta\alpha}\quad ,
\end{eqnarray}
where $X_L^{\alpha\beta}$ is the analogue of the
bosonic coordinates for the chiral basis of standard $D=4$ superspace.
The above components are the {\it complex} bosonic scalar and the complex
fermionic spinor fields
obeying the free higher spin equations in tensorial space form \cite{V01s},
\begin{eqnarray}
\label{HSpinEq=b+f}
 \partial_{\alpha [\gamma }\partial_{\delta ]\beta } \phi(X)=0\; ,
\qquad   \partial_{\alpha [\beta }\psi_{\gamma ]} (X)=0\;  . \qquad
\end{eqnarray}
Let us recall that the ${\cal N}=1$ supermultiplet contains a real bosonic scalar and
a real fermionic spinor field that obey the same equations (\ref{HSpinEq=b+f}). Hence,
the ${\cal N}=2$ supermultiplet of the conformal fields in tensorial
superspace is given by the complexification of the  ${\cal N}=1$ supermultiplet.

 Clearly, the above results are $n$-independent and thus, besides $n=4$, they
are also valid for the $n=8$ and $n=16$ cases corresponding to the $D=6$ and $D=10$
multiplets of massless conformal higher spin fields.

\subsection{${\cal N}=4$}
\label{secN4}

In contrast with the ${\cal N}=2$ case, spin-tensorial
components are present when ${\cal N}> 2$. For ${\cal N}=4$, the general
solution of the  superfield equations (\ref{*DPhi=0}) and
(\ref{asDDPhi=0}) is given by
\begin{eqnarray}
 \label{Phi=N4}
 \Phi (X,\Theta^q ,\bar{\Theta}_q)=
 \phi(X_L )+ i\Theta^{\alpha q}  \psi_{\alpha q} (X_L) +
 \epsilon_{pq}\Theta^{\alpha q}\Theta^{\beta p}  {\cal F}_{\alpha \beta} (X_L) \; ,  \qquad
 \\
\label{XL= N4} X_L^{\alpha\beta}=X^{\alpha\beta}+2i\Theta^{q(\alpha }\bar{\Theta}{}^{\beta
)}_q \; , \; q=1,2 \; ,  \qquad
\end{eqnarray}
where, again, the complex scalar and spinor fields obey the standard higher spin equations in
their tensorial superspace form,
\begin{eqnarray}
\label{HSpin4N=b+f}
  \partial_{\alpha [\gamma }\partial_{\delta ]\beta } \phi(X)=0\; ,
  \qquad   \partial_{\alpha [\beta }\psi_{\gamma ]q} (X)=0\;  , \qquad
\end{eqnarray}
while the complex symmetric bi-spinor (or `$s$-tensor' \cite{V01s}) ${\cal F}_{\alpha
\beta}={\cal F}_{ \beta\alpha}$ satisfies the tensorial counterpart of the $D=4$ Maxwell
equations (when these are written in spinorial notation \cite{Penrose}, see also below),
\begin{eqnarray}
\label{HSpin4N=cF}
\partial_{\alpha [\gamma }{\cal F}_{\delta ]\beta } (X)=0\; , \qquad
{\cal F}_{\alpha \beta}={\cal F}_{ \beta\alpha}\; . \qquad
\end{eqnarray}

However, one can easily show that the general solution of  Eq.~(\ref{HSpin4N=cF})
is expressed through a new complex scalar superfield $\tilde{\phi}(X)$
satisfying the bosonic tensorial space equation in
(\ref{HSpin4N=b+f}),
\begin{eqnarray}
\label{HSpincF=}
  {\cal F}_{\alpha \beta}= \partial_{\alpha \beta } \tilde{\phi}(X)\; , \qquad \\
\label{HSpintp=}
  \partial_{\alpha [\gamma }\partial_{\delta ]\beta } \tilde{\phi} (X)=0\; . \qquad
\end{eqnarray}

\subsubsection{$n=4\,,\, D=4$}

To prove this when $n=4$ ($\alpha,\beta =1,2,3,4$), we begin by decomposing
the complex symmetric $GL(4)$ tensor ${\cal F}_{\alpha \beta}={\cal F}_{ \beta\alpha}$
in $2\times 2$ blocks, thus keeping only the $GL(2,\bb{C})$ manifest symmetry,
\begin{eqnarray}
\label{cF=F-V-V-F}
  n=4: \quad {\cal F}_{\alpha \beta}= \left(\matrix{F_{AB} & V_{A\dot{B}} \cr
  V_{B\dot{A}} & F_{\dot{A}\dot{B}}} \right) \; , \qquad A,B=1,2\; , \quad \dot{A},
  \dot{B}=1,2 \; . \qquad
\end{eqnarray}
Let us first notice that the block components of Eq.~(\ref{HSpin4N=cF}) which contain the
antisymmetric tensors (encoded in the symmetric spin-tensors $F_{AB}$ and $
F_{\dot{A}\dot{B}}$) only,
\begin{eqnarray}
\label{dF=0=d*F}
\partial_{\dot{A}[B}F_{C]D}=0\; ,  \qquad
\partial_{{A}[\dot{B}}F_{\dot{C}]\dot{D}} =0\; ,  \qquad
\end{eqnarray}
are equivalent to the Maxwell equations for the complex selfdual field $F_{ab}={i\over
2}\epsilon_{abcd}F^{cd}\propto \sigma_{ab}{}^{CD}F_{CD}$ {\it i.e.}, they imply
$\partial^aF_{ab}=0$ and $\partial_{[a}F_{bc]}=0$.

Consider now the components of Eq.~(\ref{HSpin4N=cF}) which contain the complex vector
$V_{A\dot{B}}=\sigma^a_{A\dot{B}} V_a$ only, namely
\begin{eqnarray}
\label{dvV=}
\partial_{A [\dot{B}}V_{\dot{C}]D}=0
 \; , \qquad \partial_{\dot{B}[A}V_{D]\dot{C}}=0
 \;  \qquad
\end{eqnarray}
and
\begin{eqnarray}
\label{dtV=}
\partial_{A[B}V_{C]\dot{D}}=0
 \; , \qquad \partial_{\dot{A}[\dot{B}}V_{\dot{C}]D}=0
 \;  . \qquad
\end{eqnarray}
 Eqs.~(\ref{dvV=}) imply $\partial_{[a} V_{b]}=0$ and $\partial^aV_a=0$.
The first is solved by $V_a=\partial_{a}\tilde{\phi}$ and the second implies that the scalar
field $\tilde{\phi}$ obeys the Klein-Gordon equation $\partial^a\partial_a\tilde{\phi}=0$. In
the spin-tensor notation these read
\begin{eqnarray}
\label{V=dtp} V_{A\dot{B}}= \partial_{A\dot{B}}\tilde{\phi}\; , \qquad
\partial_{A[\dot{B}} \partial_{\dot{C}]D}\tilde{\phi}=0\; . \qquad
\end{eqnarray}

Next, the components of Eq.~(\ref{HSpin4N=cF})  which contain both vector and antisymmetric
tensor components, $\partial_{{A}\dot{B}}F_{CD}- \partial_{{A}C}V_{D\dot{B}}=0$ and
$\partial_{{A}\dot{B}}F_{\dot{C}\dot{D}}- \partial_{\dot{C}\dot{D}}V_{A\dot{B}}=0$, can be
written in the form
\begin{eqnarray}
\label{dF-ddP=}
\partial_{{A}\dot{B}}(F_{CD} - \partial_{CD}\tilde{\phi})=0\; , \qquad
\partial_{{A}\dot{B}}(F_{\dot{C}\dot{D}} - \partial_{\dot{C}\dot{D}}\tilde{\phi})=0\; ,
\qquad
\end{eqnarray}
the only covariant solution of which is given by
\begin{eqnarray}
\label{F=dP} F_{CD} = \partial_{CD}\tilde{\phi}\; , \qquad F_{\dot{C}\dot{D}} =
\partial_{\dot{C}\dot{D}}\tilde{\phi}\; . \qquad
\end{eqnarray}
Keeping in mind the Maxwell equations (\ref{dF=0=d*F}), one finds that the scalar field
$\tilde{\phi}(X)$ satisfies, besides the Klein-Gordon equation in (\ref{V=dtp}), also the
remaining components of Eq.  (\ref{HSpintp=}),
\begin{eqnarray}
\label{ddtp=0}
\partial_{A[B}\partial_{C]D}\tilde{\phi}=0\; , \qquad
\partial_{\dot{A}[\dot{B}} \partial_{\dot{C}]\dot{D}}\tilde{\phi}=0\; . \qquad
\end{eqnarray}

\subsubsection{Proof for arbitrary $n$}

We now prove that Eqs.~(\ref{HSpincF=}), (\ref{HSpintp=}) provide the general
solution of the Maxwell-like equation in tensorial space, Eq.~(\ref{HSpin4N=cF}),
for any $n$. The Fourier transform of Eq. (\ref{HSpin4N=cF}) is
\begin{eqnarray}
\label{HSpin4N=cFp}
p_{\alpha [\gamma }{\cal F}_{\delta ]\beta}(p)=0\; . \qquad
\end{eqnarray}
The solution of this equation is nontrivial {\rm iff} the matrix of the
generalized momentum has rank one, this is to say when
$p_{\alpha\beta}=\lambda_\alpha\lambda_\beta$  for arbitrary $\lambda_\alpha\not= (0,...,0)$
or, equivalently, when this matrix obeys $p_{\alpha [\gamma }p_{\delta ]\beta}=0$.
The general solution is characterized by
${\cal F}_{\alpha\beta}(\lambda) =\lambda_{\alpha}\lambda_{\beta} \phi(\lambda )$
and can be equivalently written in the form
${\cal F}_{\alpha\beta}(p) =p_{\alpha\beta} \tilde{\phi}(p)$ if
$ p_{\alpha [\gamma }p_{\delta ]\beta} \tilde{\phi}(p)=0$ \footnote{ More formally, the solution
of this equation is given by a distribution with support on
the subspace of tensorial momentum space defined by the condition
$p_{\alpha [\gamma }p_{\delta ]\beta}=0$, so that ${\tilde \phi}(p)
\propto \delta (p_{\alpha [\gamma }p_{\delta ]\beta } )$. }
of  Eqs.~(\ref{HSpin4N=cF}).
As far as set of equations
\begin{eqnarray}
\label{HSpin4NcF=ppP}
{\cal F}_{\alpha\beta}(p) =p_{\alpha\beta} \tilde{\phi}(p) \; , \qquad
p_{\alpha [\gamma }p_{\delta ]\beta} \tilde{\phi}(p)=0\;  \qquad
\end{eqnarray}
provide the Fourier transforms of Eqs.~(\ref{HSpincF=}), (\ref{HSpintp=}),
these describe the general solution.

\subsubsection{Peccei-Quinn-like symmetry}

Thus, the ${\cal N}=4$ higher spin supermultiplet actually contains two complex scalar fields
and two Dirac spinor fields in tensorial space, $\phi(X), \psi^1_\alpha (X),\psi^2_\alpha
(X), \tilde{\phi}(X)$, which satisfy the free bosonic and fermionic higher spin equations,
Eqs.~(\ref{HSpin4N=b+f}), (\ref{HSpintp=}). They appear in the on-shell scalar superfield
decomposition as
\begin{eqnarray}
\label{Phi(X)4N=} \Phi (X,\Theta^q ,\bar{\Theta}_q)= \phi(X_L )+ i\Theta^{\alpha q}
\psi_{\alpha q} (X_L) + \epsilon_{pq}\Theta^{\alpha q}\Theta^{\beta p} \partial_{\alpha
\beta} \tilde{\phi} (X_L)\; , \qquad q,p=1,2\; . \qquad
\end{eqnarray}
However, as the second complex scalar field $ \tilde{\phi}$ enters the original
superfield with a derivative, its zero mode is not fixed. In other words, this scalar is
axion-like: it possesses the Peccei-Quinn-like symmetry
\begin{eqnarray}
\label{Pe-Qu=sym4N}
 \tilde{\phi} (X)\mapsto  \tilde{\phi} (X) + const\; .
\end{eqnarray}

\subsection{${\cal N}=8$}
\label{N8part}

For higher ${\cal N}>4$ the general solution of the set of superfield equations
(\ref{*DPhi=0}) and (\ref{asDDPhi=0}) is given by
\begin{eqnarray}
\label{Phi=N>4} \Phi (X,\Theta^q ,\bar{\Theta}_q)= \phi(X_L )+ i\Theta^{\alpha q}
\psi_{\alpha q} (X_L) + \sum\limits_{k=2}^{{\cal N}/2} {1\over k!} \Theta^{\alpha_k
q_k}\ldots \Theta^{\alpha_1 q_1} {\cal F}_{\alpha_1 \ldots \alpha_k \; q_1 \ldots  q_k}
(X_L)\;  , \qquad \\ \label{F=Fsa} {\cal F}_{\alpha_1 \ldots \alpha_k \; q_1 \ldots  q_k}
(X_L)= {\cal F}_{(\alpha_1 \ldots \alpha_k ) \; [q_1 \ldots  q_k]} (X_L)\; , \;
X_L^{\alpha\beta}=X^{\alpha\beta}+2i\Theta^{q(\alpha }\bar{\Theta}{}^{\beta )}_q\; ,\;
q=1,...,4\,, \quad
\end{eqnarray}
where $\phi(X_L)$ and $\psi_{\alpha q}(X_L)$ obey the standard higher spin equations
(\ref{HSpinEq=b+f}) while the higher components satisfy
\begin{eqnarray}
\label{HSpinN>4=cF}
\partial_{\alpha [\gamma }{\cal F}_{\delta ]\beta_2\ldots \beta_k \; q_1 \ldots  q_k}
(X_L)=0\; , \qquad
{\cal F}_{\alpha_1\ldots  \alpha_q}={\cal F}_{(\alpha_1\ldots  \alpha_q)}\; . \qquad
\end{eqnarray}

   For instance, for ${\cal N}=8$ the superfield solution of the higher spin equations
(\ref{*DPhi=0}) and (\ref{asDDPhi=0})  reads
\begin{eqnarray}
\label{Phi=N8} \Phi (X,\Theta^q ,\bar{\Theta}_q)= \phi(X_L )+ i\Theta^{\alpha q}
\psi_{\alpha q} (X_L) + {1\over 2} \Theta^{\alpha_2 q_2} \Theta^{\alpha_1 q_1} {\cal
F}_{\alpha_1 \alpha_2 \; q_1 q_2}(X_L ) + \quad \nonumber \\ + {i\over 3!}\Theta^{\alpha_3
q_3}\Theta^{\alpha_2 q_2}\Theta^{\alpha_1
q_1}\epsilon_{q_1q_2q_3q}\psi_{\alpha_1\alpha_2\alpha_3}^q (X_L) + \quad \nonumber \\ +
{1\over 4!}\epsilon_{q_1q_2q_3q_4}  \Theta^{\alpha_4 q_4}\ldots \Theta^{\alpha_1 q_1}{\cal
F}_{\alpha_1 \ldots \alpha_4 }(X_L)\;  .
\end{eqnarray}
Its two lowest components obey Eqs.~(\ref{HSpinEq=b+f}), while its higher
order nonvanishing field components satisfy Eqs.~(\ref{HSpinN>4=cF}),
\begin{eqnarray}
\label{HSpinN8=cF}
\partial_{\alpha [\gamma }{\cal F}_{\delta ]\beta\; q_1q_2} (X)=0\; , \qquad \\
\label{HSpinN8=psi}
\partial_{\alpha [\gamma }{\psi}_{\delta ]\beta_2\beta_3}{}^q (X)=0  \, , \qquad \\
\label{HSpinN8=cF4}
\partial_{\alpha [\gamma }{\cal F}_{\delta ]\beta_2\beta_3\beta_4 } (X)=0\; . \qquad
\end{eqnarray}
It is tempting to identify (\ref{HSpinN8=psi}) with the tensorial space generalization of the
Rarita-Scwinger equations and Eq.~(\ref{HSpinN8=cF4}) with that of the linearized conformal
gravity equation imposed on Weyl tensor. However, similarly to the ${\cal N}=2$ case in
Sec.~\ref{Nigual2}, it is possible to show that the general solutions of Eqs.~(\ref{HSpinN8=cF}),
(\ref{HSpinN8=psi}) and (\ref{HSpinN8=cF4}) are expressed in terms of a sextuplet of
scalar fields $\phi_{q_1q_2}(X)=\phi_{[q_1q_2]}(X)$, a quadruplet of spinorial fields
$\tilde{\psi}_{\alpha_3}^q$ and  a singlet of scalar field $\tilde{\phi}(X)$ obeying
the standard tensorial space fermionic and bosonic higher spin equations
(\ref{HSpinEq=b+f}),
\begin{eqnarray}
\label{HSpin8N=sc} {\cal F}_{\alpha\beta\; q_1q_2} (X)=\partial_{\alpha \beta }
\phi_{q_1q_2}(X) \; , \qquad  \\ \label{HSpin8N=sp} \psi_{\alpha_1\alpha_2\alpha_3}^q (X)=
\partial_{\alpha_1\alpha_2} \tilde{\psi}_{\alpha_3}^q (X)  \;  , \qquad \\
\label{HSpin8N=sc4} {\cal F}_{\alpha_1 \ldots \alpha_4 } (X)=
\partial_{\alpha_1 \alpha_2 }\partial_{\alpha_3 \alpha_4 }{\tilde \phi}(X)\; . \qquad
\end{eqnarray}

Summarizing, the ${\cal N}=8$ supermultiplet of free higher spin fields is described by a set
of two scalar fields, a sextuplet of scalar fields, a spinor field and a quadruplet of
spinorial fields, all in tensorial superspace, which obey the usual type free higher spin
equations
\begin{eqnarray}
\label{HSpinEq=8N}
\partial_{\alpha [\gamma }\partial_{\delta ]\beta } \phi(X)=0\; , \qquad
\partial_{\alpha [\beta }\psi_{\gamma ]} (X)=0\;  ,  \qquad \nonumber
\\
\partial_{\alpha [\gamma }\partial_{\delta ]\beta } \phi_{qp}(X)=0\; , \qquad    \nonumber
\\
\partial_{\alpha [\beta }\tilde{\psi}_{\gamma ]}{}^q (X)=0\;  ,
\qquad \partial_{\alpha [\gamma }\partial_{\delta ]\beta } \tilde{\phi}(X)=0\; .  \qquad
\end{eqnarray}
Thus, we conclude that, in tensorial superspace, at least all the free field dynamics is
carried by the scalar and spinor fields.  However, the `higher' scalar and spinor
fields appear in the basic superfield under the action of one or two derivatives and,
hence, the model is invariant under the following generalized bosonic and
fermionic Peccei-Quinn-like symmetries
\begin{eqnarray}
\label{Pe-Qu=8Nsym}
 \phi_{qp}(X)&\mapsto & \phi_{qp}(X)+ a_{qp}\; , \qquad \nonumber \\
 \tilde{\psi}_{\alpha}{}^q (X)& \mapsto & \tilde{\psi}_{\alpha}{}^q (X)+ \beta_{\alpha}{}^q
 \; , \qquad \nonumber \\
 \tilde{\phi}(X) & \mapsto & \tilde{\phi}(X) + a + X^{\alpha\beta} a_{\alpha\beta} \;  ,
\end{eqnarray}
with constant bosonic parameters $a_{qp}=-a_{pq}$, $a$, $a_{\alpha\beta}$ and constant
fermionic parameter $\beta_{\alpha}{}^q $. Note that the non-constant shift in
$\tilde{\phi}(X)$ is allowed by the presence of two derivatives in Eq.~(\ref{HSpin8N=sc4}).

\section{Conclusion and discussion}
\label{conclu}

\medskip

In this paper we have obtained the superfield equations in ${\cal N}=2,4$ and $8$
extended tensorial superspaces $\Sigma^{(10|\mathcal{N}4)}$, which describe the
supermultiplets of the $D=4$ massless conformal free higher spin field theory with
$\mathcal{N}$-extended supersymmetry. Actually our results are valid  for
$\Sigma^{({n(n+1)\over 2}|{\cal N}n)}\,$ with arbitrary $n$ and hence for other $D$'s;
we have elaborated the  cases for ${\cal N}=2,4,8$ since for $n=4$ these
have clear `lower spin' $D=4$ counterparts  ({\it e.g.} SYM for ${\cal N}=4$,
supergravity for ${\cal N}=8$)

The ${\cal N}=2$ supermultiplet of massless conformal higher spin equations is
simply given by the complexification of the ${\cal N}=1$ supermultiplet.
For ${\cal N}=4, 8$ the $\mathcal{N}$-extended superfields contain higher components
carrying symmetric tensor  representations of $GL(n,\mathbb{R})$ that
satisfy first order equations in tensorial superspace.
It is tempting to identify these equations with a tensorial superspace
generalization of  the Maxwell equations and, in the ${\cal N}=8$ case, also with
the Rarita-Schwinger and  linearized conformal gravity equations. However,
the general solutions of these equations are  expressed through
scalar and spinor fields in tensorial space which obey the standard higher spin
equations in their tensorial space version. As these additional scalar and spinor fields of the
$\mathcal{N}$-extended supermultiplets appear in the basic superfield under derivatives, the
theory is invariant under Peccei–-Quinn--like symmetries shifting these fields.
\medskip

The superfield equations (\ref{*DPhi=0}) and (\ref{asDDPhi=0}) were found for even
${\cal N}$. It would be interesting to consider as well the case of extended tensorial
superspaces with odd ${\cal N}>1$ and to look for any specific properties of the ${\cal
N}$-extended supermultiplets of the massless conformal higher spin fields for ${\cal N}$
odd.
\medskip

Another possible direction for future study is to generalize the construction presented here
to the $OSp(\mathcal{N}|2n)$ supergroup manifolds, which provide
\cite{Frons:85,BLS99,BLSP2000,BPST04} the AdS generalization of  the tensorial
(super)spaces, and to compare the results with the free field limit of \cite{S+S=98,ESS=02}.
\bigskip

\noindent
{\it Acknowledgements.}
The authors are thankful to Sergey Kuzenko for useful comments.  This paper has been
partially supported by the research grant FIS2008-1980 from the Spanish MICINN
and by the Basque Government Research Group Grant ITT559-10.

\end{document}